\documentclass[aps, prd, superscriptaddress, nofootinbib, twocolumn, preprintnumbers]{revtex4-2}

\usepackage{bm,amssymb,slashed,graphicx,multirow,soul,mathtools,xspace,array}
\usepackage{enumitem}
\usepackage{makecell}
\usepackage{xcolor}
\usepackage{placeins}

\definecolor{nicered}{rgb}{0.7,0.1,0.1}
\definecolor{nicegreen}{rgb}{0.1,0.5,0.1}
\definecolor{niceblue}{rgb}{0.1,0.1,0.5}

\usepackage[bookmarks=false]{hyperref}
\hypersetup{colorlinks,citecolor= nicegreen,linkcolor= niceblue,urlcolor= niceblue}

\newcommand{\TeV}{\text{TeV}}

\newcommand{\MeV}{\text{MeV}}
\newcommand{\g}{\gamma}
\newcommand{\mKLrec}{m_{K_L}^{\text{rec}}}

\newcommand{\malprec}{m_{a}^{\text{rec}}}

\makeatletter
\g@addto@macro\bfseries{\boldmath}
\makeatother

\allowdisplaybreaks
\begin{document}

\title{Sweeping the pion chimney for axion-like particles with KOTO}

\author{Reuven Balkin}
\affiliation{Department of Physics, University of California Santa Cruz and Santa Cruz Institute for Particle Physics, 1156 High St., Santa Cruz, CA 95064, USA}
\author{Stefania Gori}
\affiliation{Department of Physics, University of California Santa Cruz and Santa Cruz Institute for Particle Physics, 1156 High St., Santa Cruz, CA 95064, USA}
\author{Dean J. Robinson}
\affiliation{Ernest Orlando Lawrence Berkeley National Laboratory, 
University of California, Berkeley, CA 94720, USA}
\affiliation{Berkeley Center for Theoretical Physics, 
Department of Physics,
University of California, Berkeley, CA 94720, USA}
\author{Christiane Scherb}
\affiliation{Ernest Orlando Lawrence Berkeley National Laboratory, 
University of California, Berkeley, CA 94720, USA}
\affiliation{Berkeley Center for Theoretical Physics, 
Department of Physics,
University of California, Berkeley, CA 94720, USA}

\begin{abstract}
We demonstrate that novel limits on prompt axion-like particles (ALPs) in the hard-to-probe mass range 
near the neutral pion---the so-called pion chimney---may be obtained from recasting $K_L \to 3\pi^0 \to 6\gamma$ data taken by the J-PARC KOTO experiment, to search for $K_L \to 2\pi^0a \to 6\g$.
We also explore the power of KOTO $6\gamma$ data to probe $K_L \to 2\pi^0a$ for a broader range of ALP masses, incorporating displaced decays.
\end{abstract}

\maketitle

\section{Introduction}
It is notoriously difficult to probe axion-like particles (ALP)
in a narrow interval of ALP masses nearby to the neutral pion, 
and with effective couplings $c/f_a \gtrsim 10\,\TeV^{-1}$: the so-called pion chimney.~\footnote{%
At least, so-called by S. Knapen.}
This difficulty arises for several interrelated reasons: 
(i) At the effective theory level, ALP-pion mixing is resonantly enhanced for $m_a \simeq m_{\pi^0}$.
Thus, for couplings at which the ALP is usually a long-lived state, it instead becomes prompt, 
severely limiting the sensitivity of searches in which it is assumed to be invisible in the detector acceptance,
as well as the sensitivity of direct searches for $a \to \g\g$ (i.e. with negligible electron branching ratio) 
that require some displacement of the ALP decay
in order to separate the signal from $\pi^0 \to \g\g$ backgrounds.
This results in an upper limit on the ALP couplings that can be probed.
Further, (ii) limits on prompt $a \to \g\g$ from recasts of precision diphoton measurements
are generically unable to probe ALP masses near $m_{\pi^0}$, 
because such measurements feature cuts on the diphoton invariant mass, $m_{\g\g}$,
in order to avoid being overwhelmed by $\pi^0 \to \g\g$ backgrounds.
As an example, the KTeV data for $K_L \to \pi^0 \g\g$~\cite{KTeV:2008nqz}, 
which otherwise sets powerful limits on ALP couplings, 
applies an exclusionary cut on the invariant mass window $100 < m_{\g\g} < 160$\,\MeV,
and thus no limit can be derived in that mass regime.

In this paper we show that one may leverage KOTO data for $K_L \to 3\pi^0 \to 6\g$
to obtain novel limits on $K_L \to 2\pi^0 a \to 6\g$ for prompt ALPs specifically in the pion chimney,
which we take to be $120 \le m_a \le 150\,\MeV$.
We focus on a KOTO dataset corresponding to $2\times 10^{14}$ protons on target (POT)
collected during the first run and used for the neutral flux calibration measurement,
and for which $K_L \to 3\pi^0 \to 6\g$ data is public.
In contradistinction to most other search concepts, this approach specifically targets ALPs nearby the pion mass,
because only their $K_L \to 2\pi^0 a \to 6\g$ decays are able to pass the tight experimental selections for the $K_L$ reconstruction.
We show such an ALP signal would induce distortions 
in the precisely-measured distribution of the $K_L$ reconstructed mass, $\mKLrec$.
In particular, we show the ALP signal generates single or multipeak structures that are shifted with respect to the peak at $m_{K_L}$,
and thus enable novel limits on the underlying ALP couplings to be derived.
(This approach might also be applied to the $K_L \to 2\pi^0 \to 4 \g$ flux calibration data, 
but it is likely much more challenging because of the lower statistics and $K_L \to 3\pi^0$ backgrounds, 
plus likely degradation of the reconstruction efficiency under the details of the KOTO selection.
See also Ref.~\cite{Gori:2020xvq} for a proposed $K_L \to \pi^0 a$ search at KOTO outside the pion chimney region.) 
These limits are distinct from other KOTO searches, 
such as for $K_L \to \pi^0 X$, with $X$ missing energy~\cite{KOTO:2018dsc},
or $K_L \to XX \to 4\g$~\cite{KOTO:2022lxx}.
The techniques developed in this work might also be used to derive limits from other $K_L \to 3\pi^0$ datasets, such as from KTeV~\cite{KTeV:2008gel}, provided sufficient detail regarding experimental selections and reconstruction is available to permit a recast. We leave the study of the adaptability and amenability of these techniques to other experiments for future study.

The $K_L \to 2\pi^0 a$ decay probes a similar variety of New Physics (NP) couplings as $K_L \to \pi^0 a$, 
but with differing sensitivities depending on the $P$ and/or $CP$ violating nature of the NP current.
Very close to the neutral pion mass $K_L \to 2\pi^0 a$ can provide a stronger probe of certain NP models
than $K_L \to \pi^0 a$ or even $K^+ \to \pi^+ a$,
because of suppressed $CP$-violation and weak-interaction-induced resonant mixing, respectively~\cite{Balkin:2025rqe} (see Fig.~5 therein),
Thus, constraining ALP couplings via $K_L \to 2\pi^0 a$ searches is phenomenologically well-motivated. For further discussion on the the connection of ALPs to flavor physics and flavor probes of ALPs see e.g.,~\cite{Goudzovski:2022vbt,PhysRevLett.124.071801,Izaguirre:2016dfi,Freytsis:2009ct,Bauer:2021mvw,PhysRevLett.127.081803,Bauer:2020jbp,Chala:2020wvs,MartinCamalich:2020dfe,Choi:2017gpf,CELIS2015117,PhysRevD.91.056005,PhysRevD.57.5875,Babu:1992cu,DAVIDSON1984647,PhysRevD.29.1504,REISS1982217,PhysRevLett.48.11,Ema:2016ops,Calibbi:2016hwq,Cornella:2023kjq,Cornella:2019uxs,Bjorkeroth:2018dzu,Ertas:2020xcc, Gavela:2019wzg}.

This work is structured as follows. 
In Sec.~\ref{sec:KOTOsel} we outline the KOTO detector simulation, event selection and $K_L$ reconstruction.
To simulate $K_L \to 2\pi^0 a \to 6\g$ contributions to the $\mKLrec$ spectrum, 
we develop a simplified simulation of the KOTO selection and reconstruction, 
and verify it faithfully reproduces the observed $K_L \to 3\pi^0 \to 6\g$ signal peak.
In Sec.~\ref{sec:fitlim}, 
we develop a phenomenological approach to parametrizing the lineshape profile 
of the Standard Model (SM) backgrounds in the measured $\mKLrec$ distribution,
and obtain novel limits on $\text{Br}[K_L \to 2\pi^0a]$ within the pion chimney regime.
In Sec.~\ref{sec:bych},
we proceed to explore the power of KOTO to constrain ALPs lighter than the neutral pion, 
via direct searches in the reconstructed diphoton invariant mass,
including the effects of displaced decays in flight of longer-lived ALPs.
(See Ref.~\cite{Afik:2023mhj} for a discussion of KOTO bounds on long-lived ALPs produced instead in the KOTO target.)
Sec.~\ref{sec:conc} concludes.

\section{Selection and $K_L$ reconstruction}
\label{sec:KOTOsel}
\subsection{KOTO detector system} 
\begin{figure}[tb]
	\includegraphics[width = 0.99\linewidth]{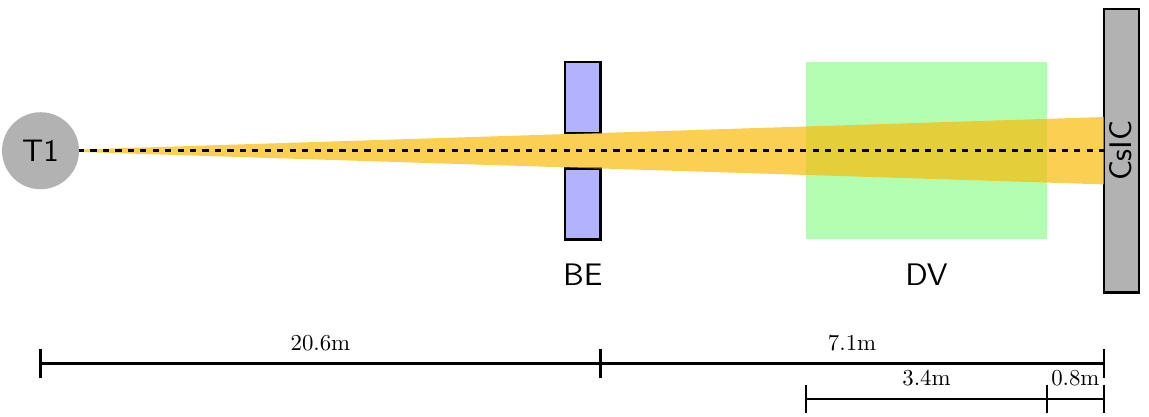}
	\caption{Schematic profile of the KOTO production target (T1), beamline (yellow), beam exit (BE), and detector system, 
	including the fiducial $K_L$ decay volume (DV) and the CsI electromagnetic calorimeter (CsIC).
	Not to scale.}
	\label{fig:kotoschem}
\end{figure}

Details of the KOTO beamline and detector system configuration can be found in Refs.~\cite{Masuda:2015eta, Masuda:2014mmc}, 
which also detail the $K_L \to 3\pi^0$ neutral flux calibration measurements that are central to the analysis in this paper (in particular, the $\mKLrec$ distribution provided in Fig.~10 of Ref.~\cite{Masuda:2015eta}).
For the purposes of this work, the salient features of the experiment are as follows: 
$K_L$'s are generated by a production target (T1) approximately 
$20.6$\,m upstream from the beam exit (BE) into the front barrel of the detector.
The fiducial $K_L$ decay volume (DV) is approximately $2.9$--$6.3$\,m downstream from the BE.
The CsI electromagnetic calorimeter (CsIC) is located approximately $7.1$\,m downstream from the BE, 
see Fig.~\ref{fig:kotoschem}.\,\footnote{%
Refs.~\cite{Masuda:2015eta, Masuda:2014mmc} use multiple overlapping coordinate systems.
The specified dimensions are based on our best understanding of the specifications therein.}
The $K_L$ momentum distribution is measured at the BE, which is an $8.5 \times 8.5$\,cm$^2$ square aperture. 
The geometric acceptance of the CsIC is a disc $1.9$\,m in diameter. 
It has a $24 \times 24$\,cm$^2$ square beam hole at its center, and makes use of two different crystal sizes:
small $25\times25$\,mm$^2$ crystals filling a square central region $1.2\times1.2$\,m$^2$ in size,
and large $50\times50$\,mm$^2$ crystals in the remaining outer region.

\subsection{KOTO detector simulation}
\label{sec:kotodetsim}

We make use of our own custom weighted Monte Carlo (MC) generator 
to emulate the core components of the KOTO detector simulation, 
including the event selection and $K_L$ mass reconstruction~\cite{Masuda:2015eta, Masuda:2014mmc}.
We define the beam line as the $z$-axis with the CsIC at $z=0$, 
such that $z_{\text{T1}} = -27.7$\,m, $z_{\text{BE}} = -7.1$\,m, and the DV lies on the interval $I_{\text{DV}} = [-4.2, -0.8]$\,m.

The $K_L$ spatial profile over the aperture is well-approximated as spherically symmetrically generated by a point source at T1.
Just as in the KOTO simulation, we generate the kaon differential flux by drawing from a parametric 
Gaussian fit to the measured $K_L$ momentum distribution at the BE~\cite{Masuda:2014mmc}.
As the energy dependence over the $K_L \to 3\pi^0$ Dalitz space varies at the (sub)percent level~\cite{ParticleDataGroup:2024cfk, KTeV:2008gel},
for the purposes of simulation we approximate the differential decay rate as pure phase space,
noting also the $\pi^0 \to 2\g$ decays proceed through an $S$-wave.
(Note the $6$-body phase space measure itself is nontrivial, 
and thus the differential rate varies over phase space.)
The MC weights are the normalized differential rates of the $K_L \to 3\pi^0 \to 6\g$ process multiplied by the $K_L$ decay and momentum probability distributions.
These are reweighted from flat prior distributions in the natural helicity basis coordinates of the $1\to 3 \to 6$-body phase space,
and flat prior $K_L$ spatial and momentum distributions.

The full KOTO detector simulation incorporates showering and clustering in the CsIC, with isolation and timing requirements.
These lead to finite energy and spatial resolution.
Based, however, on the intuition that the finite granularity of the CsIC crystals is the leading source of smearing in reconstructed observables, 
and that the showering and clustering does not bias the spatial reconstruction,
we instead approximate photon hits on the CsIC as single crystal hits, whose location is reconstructed to be at the center of the hit crystal.
We further treat the truth-level photon energy as the reconstructed energy of each hit, as the raw energy resolution of the CsIC is very high.
We will see below that these approximations lead to an excellent description of the $\mKLrec$ distribution, 
and supply a faithful representation of the KOTO reconstruction of the $K_L \to 3\pi^0 \to 6\g$ signal.
Per the KOTO selection, 
we require six reconstructed photon hits in the CsIC acceptance,
each with energy $E \ge 50$\.MeV,
and isolated by at least $150$\,mm.

The KOTO reconstruction of the $K_L$ makes use of a perturbative procedure, 
whose first step assumes the $K_L$ decay vertex is constrained to the $z$-axis, at $z=z_v < 0$.
The subsequent $\pi^0$ decays are treated as prompt.
For a pairing of hits of energy $E_i$ and $E_j$ into a putative $\pi^0$ decay, the inferred opening angle of the photon momenta 
is determined by the pion mass relation 
\begin{equation}
	\label{eqn:pimass}
	m_{\pi^0}^2 = 2E_i E_j(1 - \cos\theta_{ij})\,.
\end{equation} 
Combining $\cos\theta_{ij}$ 
with the pair's reconstructed hits on the CsIC at $\vec{r}_i$ and $\vec{r}_j$, respectively,
uniquely defines the locus of a self-intersecting torus, whose central axis lies along $\vec{d}_{ij} = \vec{r}_i - \vec{r}_j$. 
The intersection of this torus with the negative $z$-axis determines the reconstructed vertex $z_v^{ij}$ of the pair, for which
there may be $0$, $1$, or (very rarely) $2$ solutions:
for zero solutions the pairing is rejected as not reconstructing a $\pi^0$ vertex, while for two the solution with the largest $|z_v^{ij}|$ is chosen
(as in this rare case it is more probable for the $K_L$ to have decayed earlier).

There are $6!/(2^3 \times 3!) = 15$ possible sets of pairwise combinations of the six photon hits,
labelled by $\alpha=1,2, ...15$.
For each ``allowed'' set, i.e. one whose three pairs each reconstruct a vertex, 
KOTO computes a weighted average vertex
\begin{equation}
	\label{eqn:barzv}
	\bar{z}_v^\alpha = \frac{\sum_{k=1,2,3}  \, z_v^{\alpha_k}/ (\sigma_v^{\alpha_k})^2}{\sum_{k=1,2,3} 1/(\sigma_v^{\alpha_k})^2}\,,
\end{equation}
in which $\sigma_v^{\alpha_k}$ is the uncertainty of $z_v^{\alpha_k}$ 
within the reconstruction framework.
The allowed set whose weighted average has the smallest $\chi^2_\alpha = \sum_k (z_v^{\alpha_k} - \bar{z}_v^\alpha)^2 / (\sigma_v^{\alpha_k})^2$
is selected as the reconstructed photon pair combination for the event,
and its weighted average vertex is assigned to be the reconstructed $K_L$ vertex, $z_v$.
(No upper bound on $\text{min}(\chi^2_\alpha)$ appears to be imposed in the KOTO selection.)
Only events with $z_v \in I_{\text{DV}}$ pass selection. 

The second step of the KOTO reconstruction procedure perturbs the constraint that the $K_L$ vertex lies on the $z$-axis.
(Per the geometry of the square aperture at the BE, the truth-level transverse displacement of the $K_L$ vertex is small, at most $\sim 8$\,cm.)
The ``center of energy'' of the reconstructed photon hits on the CsIC, $\vec{r}_E = \sum_{i=1}^6 \vec{r}_i E_i/ \sum_i E_i$,
is used by KOTO to approximate the direction of the parent $K_L$ momentum
and thus determines the transverse displacement $\vec{r}_v$ of the parent $K_L$ vertex at $z = z_v$, 
under the assumption that the kaon was produced from a T1 point source at $z_{\text{T1}}$
using $z_v$ as determined in the first step of this procedure.~\footnote{%
The correct quantity for this is the center of longitudinal momentum, but this is not directly reconstructible at KOTO.} 
Using $\{\vec{r}_v, z_v\}$ as the $K_L$ vertex location, 
one may then reconstruct all photon momenta, and thus the $K_L$ invariant mass (as well as the pairwise pion invariant masses).

In our MC we apply the same selection and reconstruction procedure, but with the following simplifications:
based on the intuition that the $\sigma_v^{ij}$ uncertainty is mainly driven by detector CsIC crystal granularity,
which in turn scales with the areas of CsIC crystals,
we assume that all pairs with both hits on the central (small crystal) region have the same vertex uncertainty $\sigma_{v0}$. 
For pairs with one or two hits in the outer (large crystal) region, 
we then scale the uncertainty by a factor of $4$ and $16$, respectively,
per the relative areas of the small and large crystals.
As $\bar{z}_v^\alpha$ in Eq.~\eqref{eqn:barzv} and the relative ratios of $\chi^2_\alpha$ 
are unaffected by the overall scaling of the uncertainties,
under this approximation we may compute each weighted average vertex and identify $z_v$ from $\text{min}(\chi^2_\alpha)$
using the relative uncertainties alone, without needing to estimate $\sigma_{v0}$ itself.

\begin{figure}[t]
	\includegraphics[width = 0.99\linewidth]{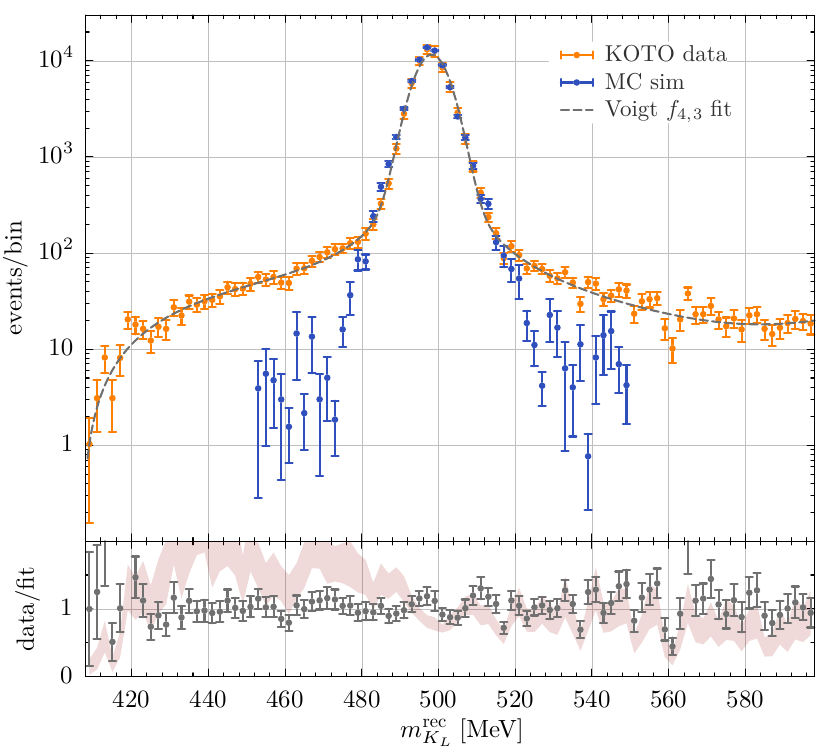}
	\caption{\textbf{Top}: KOTO data (orange)  for the reconstructed $K_L$ mass, 
	$\mKLrec$, from the $K_L \to 3\pi^0 \to 6\g$ dataset for $2 \times 10^{14}$ POT, versus our MC simulation (blue). 
	Uncertainties in the latter are due to MC statistics only. 
	Also shown is the lineshape SM background fit using the $f_{4,3}$ Voigt-like model; see Sec.~\ref{sec:linemod}.
	\textbf{Bottom}: Ratio of data versus the $f_{4,3}$ fit (dark gray points),
	and the 68\% CL band for the ratio of data versus an approximate simulation of the $K_L \to 3\pi^0$ peak 
	plus continuum contributions (light red band);
	see Sec.~\ref{sec:background}.}
	\label{fig:KOTOrec}
\end{figure}

In Fig.~\ref{fig:KOTOrec} we show the resulting $\mKLrec$ distribution (blue)
from a MC sample with $10^7$ initial $K_L$'s at the BE,
compared to KOTO data (orange) from the $2 \times 10^{14}$ POT dataset (Fig.~10 of Ref.~\cite{Masuda:2015eta}).
The simulated distribution is normalized to the total number of events in the KOTO $\mKLrec$ data---approximately $70 \times 10^3$ post-selection---by 
applying an overall scaling,
noting that the MC statistical errors over the simulated peak are negligible compared to those in the measured $\mKLrec$ distribution.
The excellent agreement over the peak verifies that our MC simulation 
provides a faithful representation of the KOTO reconstruction of the $K_L \to 3\pi^0 \to 6\g$ signal.
The long tails in the KOTO data predominantly arise from other background activity in the detector, 
such as halo neutron scattering, 
coinciding with $K_L \to 3\pi^0$ decays in which one or more photon pairs miss the calorimeter acceptance.
This ``accidental overlay'' results in six-photon-cluster events that generate a continuum background in the $\mKLrec$ distribution.
We need not simulate these tails to characterize the $3\pi^0$ signal peak, nor that of $2\pi^0a$ considered below.
Simulation of such backgrounds will be relevant in Sec.~\ref{sec:bych}, 
that considers a modified KOTO selection to search for displaced ALP decays with masses outside the pion chimney.

The acceptance in our simulation is approximately $19 \times 10^{-4}$
which is somewhat larger than the quoted $4.3\times 10^{-4}$ acceptance in the full KOTO simulation~\cite{Masuda:2015eta, Masuda:2014mmc}.
The latter incorporates additional selections as well as the abovementioned showering, clustering and timing effects, 
which may account for the difference.
The results in Fig.~\ref{fig:KOTOrec} demonstrate that these effects do not appear 
to significantly bias the shape of the $\mKLrec$ peak distribution from $K_L \to 3\pi^0 \to 6\g$.

\begin{figure*}[htb!]
	\includegraphics[width = 0.75\linewidth]{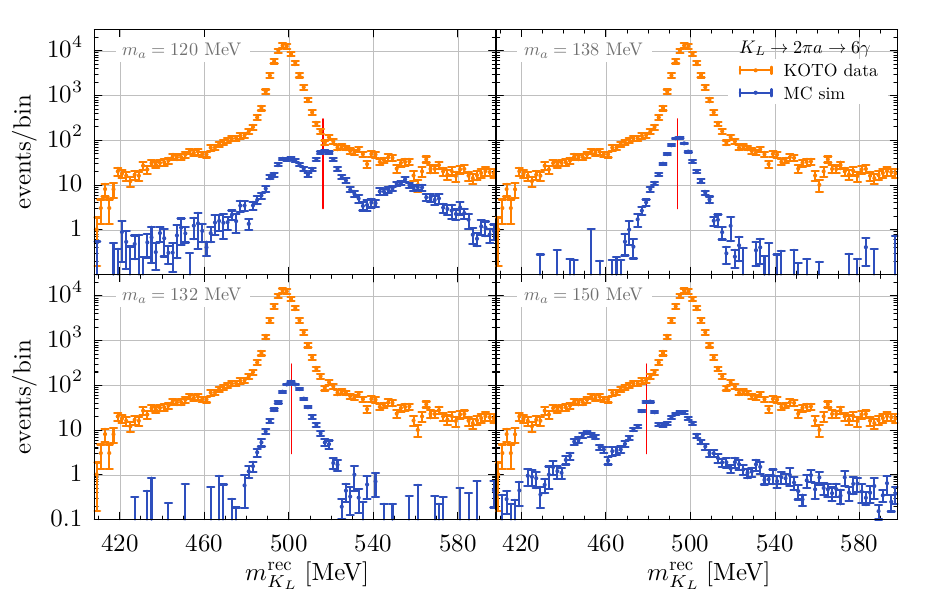}
	\caption{Simulated distributions for the reconstructed $K_L$ mass, $\mKLrec$, 
	from $K_L \to 2\pi^0 a \to 6\g$ under our emulation of the KOTO selection and reconstruction (in blue), 
	compared to the KOTO $K_L\to 3\pi^0\to6\gamma$ data (in orange) for $2\times 10^{14}$ POT.
	For visual clarity, the ALP signal distributions are normalized such that $\text{Br}[K_L \to 2\pi^0 a]/\text{Br}[K_L \to 3\pi^0] = 0.01$.
	The red vertical line indicates the expected location of the shifted $m^{\text{rec}}_{K_L}$ peak from Eq.~\eqref{eqn:deltamk}.}	
	\label{fig:KOTOrec2PiA}
\end{figure*}

\subsection{$K_L \to 2\pi^0 a$ signal}
We simulate the $K_L \to 2\pi^0 a \to 6\g$ signal by varying the mass of one pseudoscalar in our treatment of $K_L \to 3\pi^0$. 
We treat the chimney range as a $\pm 15$\,MeV window around $m_{\pi^0}$
\begin{equation}
	120~\text{MeV} \le m_a \le 150~\text{MeV}\,
\end{equation}
We use the basis-independent chiral perturbation theory (ChPT) description of the ALP production and decay amplitudes from Ref. \cite{Balkin:2025rqe} throughout.
In the chimney regime the $K_L \to 2\pi^0a$ amplitude is dominated by ALP-pion mixing,
so that corrections to the energy dependence of the $2\pi^0 a$ versus $3\pi^0$ Dalitz spaces are subleading, 
being estimated as $\lesssim 10\%$ \cite{Balkin:2025rqe},
and may be safely neglected.
Similarly, we treat the ALP as prompt over the entire chimney, 
and assume that the $a \to \g\g$ branching ratio dominates those of other decay modes,
as generically expected.
Some of our projected limits below extend to regions in which the ALP lifetime $c\tau_a\sim \text{mm}-\text{cm}$. 
This remains smaller than our estimate of the KOTO vertex resolution, which we find to be $\mathcal{O}(10\text{cm})$ by propagating the KOTO photon energy and hit resolutions (Eqs~(6.1) and~(6.2) of Ref.~\cite{Masuda:2014mmc})
through our simulation of the KOTO reconstruction.
Thus the ALPs may still be treated as prompt.

The $K_L \to 2\pi^0 a$ signal is generated with the same overall scaling as derived for the $K_L \to 3\pi^0$ sample, 
so that the ALP signal distributions are normalized to the $K_L \to 3\pi^0$ branching ratio.
We generate simulated samples with $3\times 10^6$ initial $K_L$'s at the BE---sufficient 
for subleading statistical uncertainties versus the KOTO data for $2 \times 10^{14}$ POT--- for 
ALP masses in $2$~MeV intervals over the chimney range, as well as for $m_a = m_{\pi^0}$.
For each $m_a$, the signal event sample is passed through 
our simulation of the KOTO selection and $K_L$ mass reconstruction 
to generate a $\mKLrec$ distribution.
Critically, this procedure applies the pion mass constraint~\eqref{eqn:pimass} to all pairs of photons, 
even those originating from the ALP.

Some intuition is helpful for anticipating the morphology of the resulting ALP signal lineshapes in the $\mKLrec$ distribution.
In general, for a lighter (heavier) ALP, one expects photon pairs to be produced in a typically narrower (broader) cone versus the pion.
Under a reconstruction which assumes the ALP is a pion,
one should expect the variation in the cone to appear to be the result of a typically more (less) energetic pion, 
and thus a typically heavier (lighter) reconstructed kaon.
This results in a peak that is shifted upwards (downwards) with respect to $m_{K_L}$.

We may approximately quantify how this expectation manifests through the application of Eq.~\eqref{eqn:pimass}
by taking the small opening angle limit 
and constraining the decay vertices to the $z$-axis for simplicity.
It follows from Eq.~\eqref{eqn:pimass} that for the $a \to \g\g$ vertex, 
the variation induced in the reconstructed decay vertex, $z_v^a$, and in the reconstructed pairwise opening angle, $\theta$, by the shift $\delta m = m_a - m_{\pi^0}$,
obeys $\delta z_v^a/z_v^a \simeq -\delta\theta /\theta \simeq \delta m/m_{\pi^0}$.
Further assuming the simplified case that the three reconstructed vertices--- i.e. 
$z_v^a$, $z_v^{1,2}$ for the ALP and two $\pi^0$'s, respectively---have 
equal uncertainties and that the combinatoric set corresponding to $\text{min}(\chi_\alpha^2)$ remains the same,
then misreconstruction of the ALP vertex distorts the weighted average such that $\delta \bar{z}_v = \delta z_v^a/3$.
As $z_v^a \simeq z_v^{1,2}$ up to corrections typically parametrically smaller than $\delta z_v^a$, 
then also $\bar{z}_v \simeq z_v^a$ and so
\begin{equation}
	\label{eqn:dzv}
	\delta \bar{z}_v/\bar{z}_v \simeq \delta z_v^a/3z_v^a \simeq \delta m/3m_{\pi^0}\,.
\end{equation}
Because the $K_L$ momentum is aligned with the beamline under the assumption that the decay is constrained to the $z$-axis, 
then the sum of photons' transverse momentum is zero,
while their total longitudinal momentum $\sum_i k_{L,i} \simeq \sum_i E_i(1 - r_i^2/(2 \bar{z}_v^2))$,
in which $r_i$ is the hit location of the $i$th photon on the CsIC and we take the limit $r_i \ll \bar{z}_v$.
Hence the reconstructed mass ${\mKLrec}^2 \simeq \sum_{i,j} E_i E_j r_j^2/\bar{z}_v^2$, whence $\delta \mKLrec/\mKLrec \simeq -\delta \bar{z}_v/\bar{z}_v$.
Combining this with Eq.~\eqref{eqn:dzv},
one expects that the application of Eq.~\eqref{eqn:pimass} 
should result in shifts in the peak in the $\mKLrec$ distribution, such that
\begin{equation}
	\label{eqn:deltamk}
	\delta \mKLrec \simeq  -\big[m_{K_L}/3m_{\pi^0}\big]\delta m \simeq -1.2\, \delta m\,.
\end{equation}
Beyond the small perturbation regime, one may expect combinatoric misreconstructions to play a role, 
leading to another peak near $m_{K_L}$, or elsewhere.

In Fig.~\ref{fig:KOTOrec2PiA} we show the $\mKLrec$ distributions arising from a $K_L \to 2\pi^0 a \to 6\g$ signal (blue),
versus the KOTO data (orange), for the four cases $m_a = 120$, $132$, $138$, and $150$\,MeV. 
In each case, there is a shifted peak matching the predictions from Eq.~\eqref{eqn:deltamk} (red vertical lines).
For ALP masses near the edges of the considered mass window, 
peaks at $m_{K_L}$ and in the far tails also arise, leading to a multipeak morphology for the ALP signal lineshape.

\section{Fits and Limits}
\label{sec:fitlim}
\subsection{Lineshape model}
\label{sec:linemod}
The measured $\mKLrec$ spectrum includes the $K_L \to 3\pi^0$ peak 
as well as a continuum contribution that results in the extended high and low tails. 
To set limits on the $K_L \to 2\pi^0 a$ contribution from data requires comparison to a prior 
for these SM backgrounds from either theory and/or Monte Carlo,
in order to obtain a background-subtracted dataset.
No first principles calculation of the $\mKLrec$ spectrum is likely possible.
Further, one would require a precision MC simulation 
that is self-consistently implemented to avoid tunings to other measurements that may in principle
be sensitive to the same ALP phenomenology being probed.

To avoid this, we instead adopt a phenomenological approach to obtain a lineshape description of the collective SM background.
Based on the assumption that the true backgrounds should be smooth, 
we employ a smooth functional form in terms of $\mKLrec \equiv \mu$ (for ease of reading) 
that parametrizes the expected morphology: 
we attenuate resulting model dependence via information theoretic techniques, described below.
In particular, we choose a form that incorporates the expected $3\pi^0$ kinematic threshold singularity plus a Voigt profile description of the peak,
modified by rational polynomial prefactors that vanish at $\mu = 3m_{\pi^0}$,
and have simple large $\mu$ behavior.~\footnote{%
The Voigt profile~\cite{Voigt1912}, a convolution of a Lorentzian and Gaussian, 
is very widely-used as a descriptor of an experimentally-smeared resonance peak.
It is often combined in experimental particle physics with ad hoc descriptions of continuum background distributions. 
Here we use the nonrelativistic version, though relativistic versions can also be contemplated, see. e.g. Ref.~\cite{Kycia:2017gjn}.
}
Explicitly, we consider a lineshape
\begin{multline}
	f_{2n_a,n_b}(\mu) = \sqrt{\mu^2-9m_{\pi^0}^2}\bigg\{\sum_{k=1}^{n_a} a_{2k} \frac{\big[\mu^2 - 9m^2_{\pi^0}\big]^{2k-2}}{\mu^{2k}}  \\ 
	\quad + \bigg[\sum_{k=0}^{n_b}b_k \frac{\big[\mu^2 - 9m^2_{\pi^0}\big]^{k}}{\mu^{k}}\bigg] V_{\delta, \,\sigma}\big(\mu -m_{K_L} + \delta m\big)\bigg\}\,,
	\label{eqn:linesh}
\end{multline}
in which $V_{\delta, \sigma}(\mu)$ is the Voigt profile distribution, 
proportional to the real part of $e^{-\zeta^2} \text{erfc}[i \zeta]$, $\zeta = (\mu - i \delta)/(\sqrt{2} \sigma)$.
For the purpose of numerical stability in the fits, we take starting values of $\delta = 1\,\MeV$ and $\sigma = 4\,\MeV$ in fits to Eq.~\eqref{eqn:linesh}.
The $a_{2k}$ and $b_k$ terms have large-$\mu$ dependence 
scaling as $\mu^{2k - 3}$ and $\mu^{k +1}$, respectively.
The free parameters of this model are the three Voigt parameters $\delta$, $\sigma$ and $\delta m$, 
which control the height, spread, and maximum of the peak, respectively,
as well as the rational polynomial prefactor parameters $a_{2k}$, $k = 1,\ldots,n_a$, and $b_k$, $k = 0,\ldots,n_b$.

In all fits we consider, we assume that bin-bin correlations in the $\mKLrec$ spectrum are negligible:
assuming some degree of (anti)correlation will typically weaken (strengthen) the power of the limits we derive.
Thus one should interpret the limits as prospective, 
not only because of our phenomenological description of the assumed smooth lineshape,
but also pending a fully-fledged analysis within an experimental framework that incorporates effects of correlations.
In order to avoid over or underfitting we use information theoretic techniques to select the appropriate number of the polynomial prefactors.
Considering all possible models with $n_a \le 3$ and $n_b \le 3$, in order to keep the complexity somewhat low, 
the Akaike Information Criterion (AIC)~(see e.g. Ref.~\cite{burnham2002model}) selects $f_{4,3}$, with $9$ parameters:
$a_{2,4}$, $b_{0,1,2,3}$, $\delta$, $\sigma$ and $\delta m$.
The resulting `SM background fit' is shown in Fig.~\ref{fig:KOTOrec} along with the ratio of the data versus the fit in the bottom panel.
The fit $\chi^2/\text{dof} = 124.3/(95-9) = 1.45$.

\subsection{Approach}
Caution and care must be taken in treating the residuals of the SM background lineshape fit as background-subtracted data 
to which a $K_L \to 2\pi^0a$ signal distribution can be compared, 
because of several complications.
First, just as for the SM backgrounds, there is no closed-form theory description for the multipeak morphology of the ALP signal itself,
and it cannot be simply modeled as a narrow peak on top of the fit residuals.
Second, the chosen model of the lineshape may not be fully orthogonal to the morphology of the ALP signal,
such that fits to the model may in principle absorb some of the signal lineshape features, leading to weaker limits.
That is, the limits may be overestimated if this effect is not included.

These complications motivate a combined signal plus background fit approach, as follows:
We introduce an additional parameter 
\begin{equation}
	\label{eqn:rhodef}
	\rho = \text{Br}[K_L \to 2\pi^0a]/\text{Br}[K_L \to 3\pi^0] \ge 0\,,
\end{equation}
For a given ALP mass and $\rho$ hypothesis, 
we subtract the simulated ALP signal from the measured data 
and then reperform the fit to the $f_{4,3}$ lineshape template,
yielding a ``subtracted'' fit statistic $\chi^2_{\text{sub}}(m_a, \rho)$.
Constructing a likelihood ratio with respect to the SM background fit, 
which has $\chi^2_0 = \chi^2_{\text{sub}}(m_a, 0)$, yields
\begin{equation}
	\delta\chi^2(m_a,\rho) =   \chi^2_{\text{sub}}(m_a, \rho) - \chi^2_0\,,
\end{equation}
which is strictly positive provided that the data currently sets only limits on $\rho$ (verified below),
equivalent to the hypothesis that the true value is at $\rho = 0$.
Because $\rho \ge 0$ only, 
this $\delta\chi^2$ is then distributed as a $1:1$ admixture of a zero and one degree of freedom $\chi^2$~\cite{Chernoff:1954},
 i.e. with PDF $p(x) = 1/2\big[\delta(x) + p_{\chi^2_1}(x)\big]$.
Thus a 90\% CL on $\rho$ corresponds to $\delta\chi^2(m_a, \rho) < 1.64$.

Although we have argued the residuals of the SM background fit in Sec.~\ref{sec:linemod} 
should not be used directly to obtain a limit, 
we use them to obtain a ``naive'' limit for $\rho$.
Over a range of $\rho$ values 
covering this naive limit, 
we then refit the $f_{4,3}$ lineshape to determine $\delta\chi^2(m_a, \rho)$,
and interpolate to obtain the 90\% CL.
(In principle, one should reperform the AIC model selection for each refit.
However, the shift in the criterion value from the naive to full treatment is smaller than the shift in the AIC to other models,
such that $f_{4,3}$ remains the optimal model in the considered model space.)

\subsection{Results}

\begin{figure}[tb]
	\includegraphics[width = 0.99\linewidth]{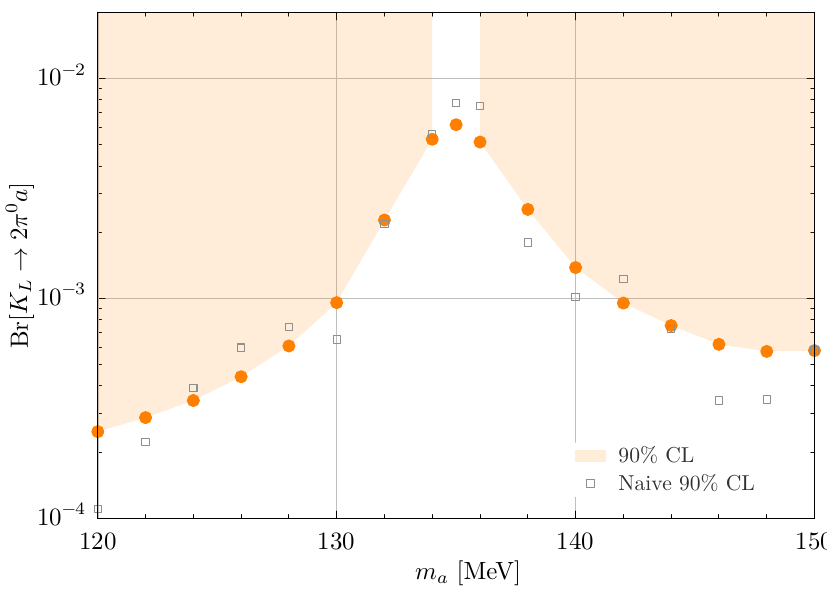}
	\caption{Interpolated limits (orange) on $\text{Br}[K_L \to 2\pi^0 a]$ at $90\%$ CL from the KOTO measurement of $m^{\text{rec}}_{K_L}$ in $K_L \to 3\pi^0 \to 6\g$,
	with $2 \times 10^{14}$ POT.
	Limits evaluated at specific $m_a$ values are shown by orange data points.
	The limit at $m_a = m_{\pi^0}$ is shown for reference, but no limit is inferred within $\pm1$\,MeV of $m_{\pi^0}$; see text.
	Also shown are naive limits at $90\%$ CL (gray squares) derived from the residuals of the SM background fit.}
	\label{fig:BrLim}
\end{figure}

\begin{figure}[tb]
	\includegraphics[width = \linewidth]{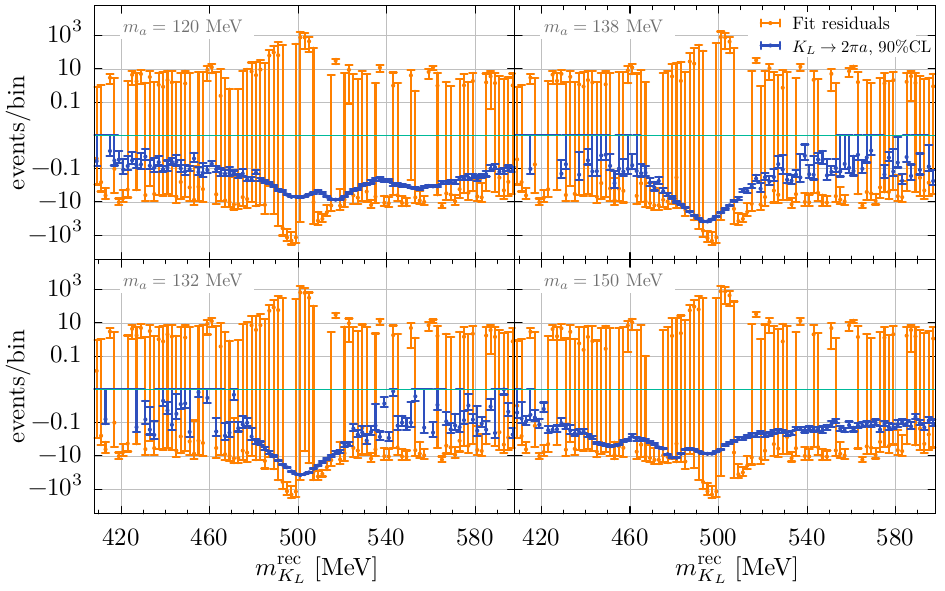}
	\caption{
	The fit residuals (orange) for a fit of the $f_{4,3}$ lineshape to the KOTO $\mKLrec$ data with subtracted ALP signal component (blue),
	evaluated at the derived limit for the branching ratio at $90\%$ CL from Fig.~\ref{fig:BrLim},
	for the same four ALP masses as in Fig.~\ref{fig:KOTOrec2PiA}.
	Fit residuals may be of either sign, or may have uncertainty ranges that fluctuate through zero (in either direction).
	We therefore show signed logarithmic plots, 
	with the logarithmic ranges soldered at $\pm10^{-3}$ events/bin, indicated by the green line.
	}
	\label{fig:KOTOrecbrlim}
\end{figure}

\begin{figure*}[htb]
    \includegraphics[width=0.55\linewidth]{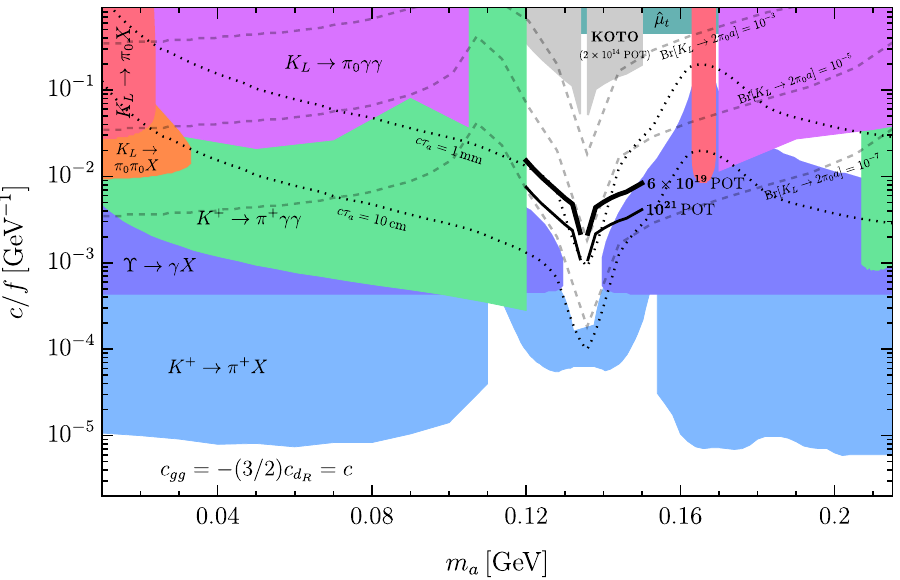} \hspace*{1cm} \\
    \includegraphics[width=0.45\linewidth]{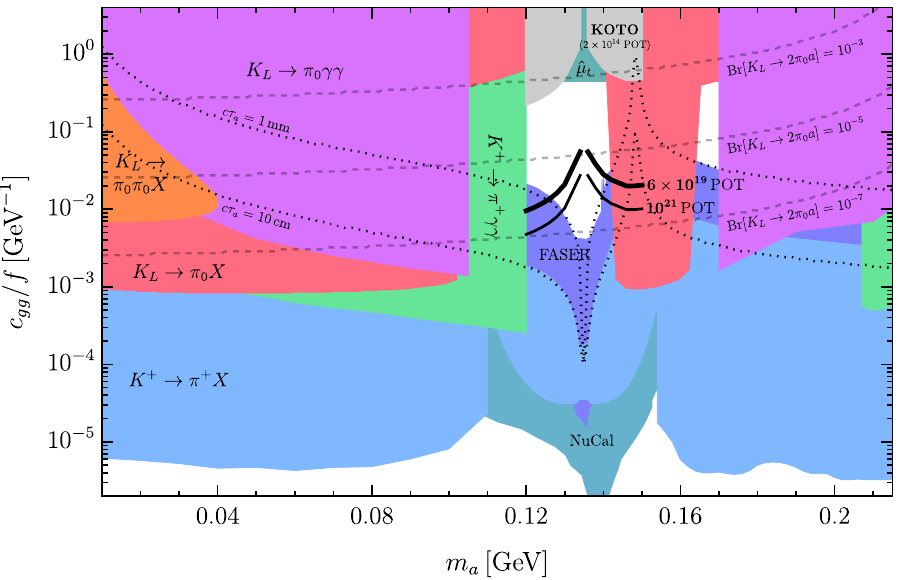}
      \includegraphics[width=0.45\linewidth]{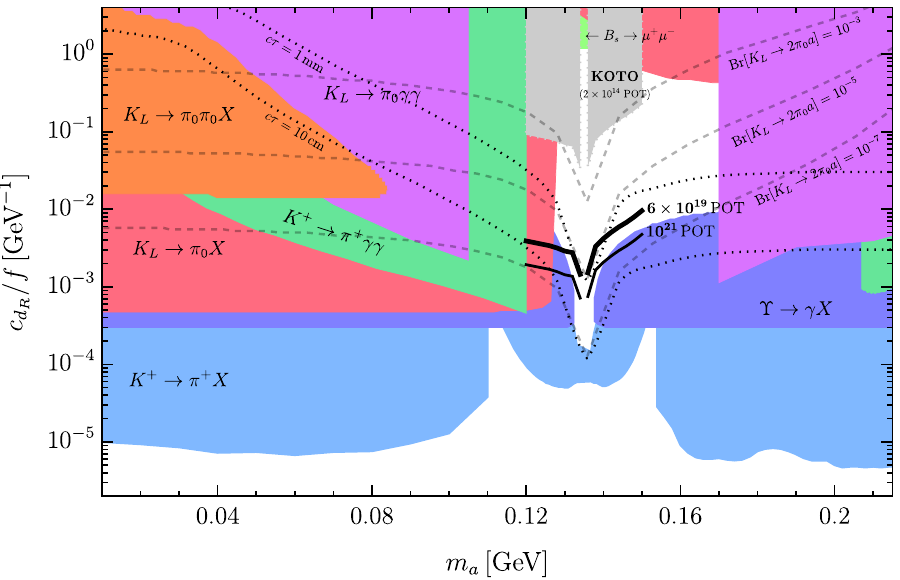}
    \caption{Prospective 90\% CL bounds on ALP couplings in the pion chimney from this analysis (gray shaded) 
     for $c_{gg}=-3/2c_{d_R}=c$ (top), $c_{gg}$ (bottom left) and $c_{d_R}$ (bottom right), per Ref.~\cite{Balkin:2025rqe}. Previous bounds include limits from $K^+\to \pi^+X$ (blue)~\cite{NA62:2021zjw}, $K^+\to\pi^+\gamma\gamma$ (green)~\cite{NA62:2023olg,E949:2005qiy}, $K_L\to\pi^0X$ (red)~\cite{KOTO:2024zbl}, $K_L\to\pi^0\gamma\gamma$ (purple)~\cite{NA48:2002xke}, $K_L\to\pi^0\pi^0X$ (orange)~\cite{E391a:2011aa}, $B_s\to \mu^+\mu^-$ (light green)~\cite{CMS:2020rox}, FASER (dark purple)~\cite{FASER:2024bbl}, NuCal (teal)~\cite{Blumlein:1991xh}, and the top chromomagnetic dipole moment $\hat{\mu}_t$ (dark green)~\cite{CMS:2019nrx} using the expression in Ref.~\cite{Bauer:2021mvw}. Details on the calculation of the bounds can be found in~\cite{Balkin:2025rqe}.
   The heavy  and light solid black lines represent our naive projections for $6\times 10^{19}$ and $10^{21}$ POT, respectively.
   A fully-fledged analysis within an experimental framework is required to validate the precise CLs.}
\label{fig:KOTO_bounds_combined}
\end{figure*}

As above, we consider ALPs with masses in the pion chimney mass range, in steps of $2$ MeV.
Noting the $2$\,MeV binning of the KOTO  $\mKLrec$ data, 
and with reference to Eq.~\eqref{eqn:deltamk},
we do not set limits within a $\pm1$\,MeV window around $m_a = m_{\pi^0}$.
Within this window $\delta\mKLrec$ is below the binning resolution, 
so that one may then expect that the lineshape fit becomes mainly sensitive to higher-order details of the shape of the simulated $\mKLrec$ peak,
which we expect is less well-controlled in our approximation of the detector simulation 
versus the location of the peak itself.
When interpreting our results in the context of limits on Wilson coefficients of particular models (see Sec.~\ref{sec:projlim}), we similarly exclude a $\sim 2$\,MeV window around $m_{\pi^0}$ because of the breakdown of the ChPT description.

In Fig.~\ref{fig:BrLim} we show the resulting limits on $\text{Br}[K_L \to 2\pi^0a] (= \rho \times \text{Br}[K_L \to 3\pi^0])$ at $90\%$ CL, 
using the known value $\text{Br}[K_L \to 3\pi^0] = 0.1952(12)$~\cite{ParticleDataGroup:2024cfk}.
For $m_a$  close to $m_{\pi^0}$, such that the ALP signal distribution is a single shifted peak, 
the naive limit (in gray) for $\rho$ coincides reasonably closely with the interpolated value from the refit treatment (in orange). 
By contrast, near the boundaries of the chimney, the interpolated limit can sometimes be $\mathcal{O}(1)$ weaker.
These results suggest that any model dependence from our choice of lineshape model for the backgrounds is at most an $\mathcal{O}(1)$ effect,
and that the limits are reliable at that level.  
From the figure, limits on branching ratios as small as $\text{few} \times 10^{-4}$ may be extracted already, 
using a luminosity as small as $2\times 10^{14}$ POT.

To gain a sense of the size of the ALP branching ratio limit versus the fit residuals, 
we show in Fig.~\ref{fig:KOTOrecbrlim} the residuals of the $f_{4,3}$ lineshape fit to the KOTO $\mKLrec$ data,
after subtracting the ALP signal component evaluated at the $90\%$ CL branching ratio limit or four different ALP mass values. 
As expected, the subtracted ALP component is comparable in size to and closely bounded by the fit residuals.

\subsection{Projected limits}
\label{sec:projlim}

The branching ratio limits may be reinterpreted in the context of a particular set of ALP-SM Wilson coefficients, 
from which the $K_L \to 2\pi^0 a$ branching ratio may be computed.
Per the analysis and conventions of Ref.~\cite{Balkin:2025rqe},
we consider three scenarios, in which the non-zero Wilson coefficients are: 
(i) $c = c_{gg} = -(3/2) c_{d_R}$, which exhibits a large enhancement of the $K_L \to 2\pi a$ rates versus $K_L \to \pi a$,
(ii) $c_{gg}$, and
(iii) $c_{d_R}$.
The latter two scenarios are commonly examined in ALP literature.
The resulting limits are shown by gray shaded regions in Figs~\ref{fig:KOTO_bounds_combined},
compared to various other known limits from NA62~\cite{NA62:2021zjw,NA62:2023olg}, NA48~\cite{NA48:2002xke}, E391A~\cite{E391a:2011aa}, BarBar~\cite{BaBar:2010eww}, E949~\cite{E949:2005qiy}, KOTO~\cite{KOTO:2024zbl} and CMS~\cite{CMS:2019nrx}: for a discussion see Ref.~\cite{Balkin:2025rqe}.
While we emphasize again that one should interpret these limits as prospective, 
we see from our analysis that the KOTO calibration data  ($2\times 10^{14}$ POT) can prospectively set novel limits in the pion chimney for all three scenarios.
If this technique is not systematically limited for the larger datasets 
comparable to the sizes of the 2016-2021 runs (roughly $6\times 10^{19}$ POT) 
or the projected full run ($10^{21}$ POT),
and assuming the reach scales as $1/\sqrt{\text{POT}}$,
then one sees from the solid black lines in Figs~\ref{fig:KOTO_bounds_combined} 
that one may probe most  of the pion chimney.

The validity of the ChPT description for the ALP interactions around $m_a\sim m_{\pi^0}$ in Fig.~\ref{fig:KOTO_bounds_combined} holds in the small angle approximation, 
\begin{equation}
     \theta_{\pi a}\sim f_\pi/f \times c_X/|\delta_{m_a}| \ll 1\,,
\end{equation}
where $c_X$ is the Wilson coefficient, $\theta_{\pi a}$ is the mixing angle between $\pi^0$ and the ALP, 
and the tuning parameter $\delta_{m_a} \equiv {m_a^2}/{m_{\pi^0}^2}-1$.
For example, even for the moderately low UV scale $c_X/f=0.1\,\text{GeV}^{-1}$ -- corresponding to the typical best reach of the limits from the $2 \times 10^{14}$ POT KOTO calibration dataset -- a few percent-level mass tuning $\delta_{m_a}\sim 0.03$ falls within the small-angle approximation, with $\theta_{\pi a} \sim 0.3$. 
This corresponds to excluding $|m_a - m_\pi| \lesssim 2$ MeV from the regime of validity, which is similar to the already-excluded interval shown on our Figs.~\ref{fig:BrLim} and~\ref{fig:KOTO_bounds_combined},
arising from finite binning resolution.
There is, in addition, a constraint on $\theta_{a\pi}$ from the $\pi^0$ lifetime (see also e.g.~\cite{Ovchynnikov:2025gpx})
which is modified by a $\cos\theta_{a\pi}^2$ factor
in the small angle regime.
Comparing the next-to-leading order ChPT prediction~\cite{Goity:2002nn} 
to the measured value~\cite{ParticleDataGroup:2024cfk} implies similarly $\theta_{a \pi} < 0.27$ at $2\sigma$~CL 
(the measured value itself exceeds the prediction at just over $2\sigma$ CL, favoring some degree of mixing.)

\begin{figure}[tb]
	\includegraphics[width = 0.95\linewidth]{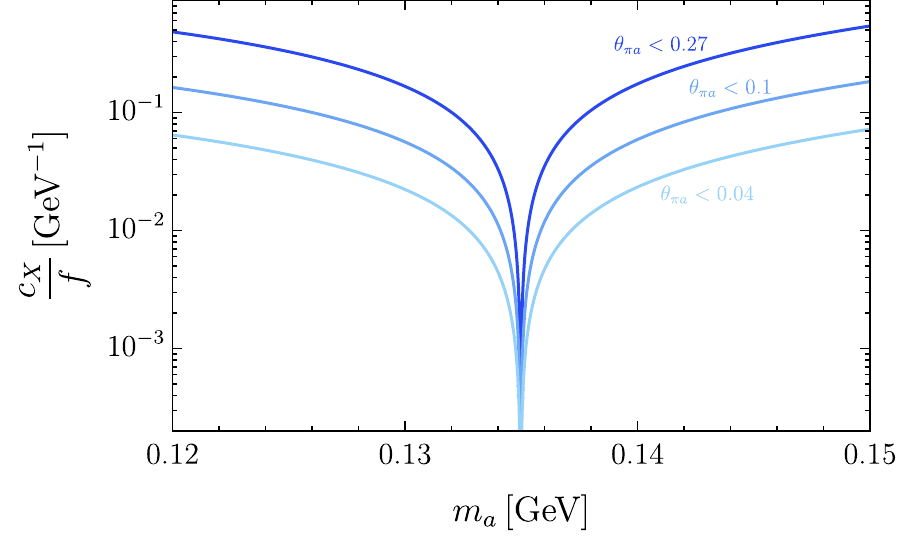}
	\caption{Upper bounds for the validity of the ChPT description of the ALP interactions for $\theta_{\pi a} < 0.1$ and $\theta_{\pi a} < 0.3$.}
	\label{fig:validity}
\end{figure}

More generally, the prospective limits on $c_X/f$ from our analysis are substantially model dependent. 
We show in Fig.~\ref{fig:validity} example upper bounds for which the ChPT description is expected to remain valid for $\theta_{\pi a} < 0.04$, $\theta_{\pi a} < 0.1$  and $\theta_{\pi a} < 0.27$.
In some models, the prospective limits for the 
$2 \times 10^{14}$ POT KOTO calibration dataset may be close to or exceed the upper bounds outside the $134~\text{MeV}\le m_a\le 136$~MeV window, depending on $\theta_{\pi a}$.
However, the projected stronger limits on $c_X/f$ (black lines in Fig.~\ref{fig:KOTO_bounds_combined})
that may be obtained from more recent or future, larger KOTO datasets, 
will fall well within the regime of validity of ChPT up to very narrow mass windows around $m_{\pi^0}$.

\section{Beyond the chimney}
\label{sec:bych}
So far we have considered limits for $K_L \to 2\pi^0 a$ signals 
that are able to misreconstruct as $K_L \to 3\pi^0$ under the KOTO selection described in Sec.~\ref{sec:kotodetsim},
which imposes Eq.~\eqref{eqn:pimass} on all photon pairs.
One may consider, however, an adaptation of the KOTO reconstruction for ALPs 
in which Eq.~\eqref{eqn:pimass} is relaxed for one photon pair,
and one instead sets limits on the $K_L \to 2\pi^0a$ branching ratio via `bump hunting' 
in the spectrum of the measured invariant diphoton mass, $\malprec$.
This technique may set more powerful ALP limits for $m_a$ far outside the chimney
compared to lineshape fits to the $\mKLrec$ spectrum
(though any such limits will currently not be competitive with constraints from other $K \to \pi a$ searches). 

\subsection{Modified reconstruction}

For $m_a$ well below $m_{\pi^0}$, it is no longer natural to treat the ALP as prompt,
and thus one must also generalize the KOTO reconstruction technique to accommodate displaced-in-flight ALP decays.
One may also consider the case of $m_a > m_{\pi^0}$ with small enough coupling that the ALP is long-lived.
In all that follows we will consider only $m_a \le m_{\pi^0}$, 
keeping in mind the proposed approach may also be applied to larger ALP masses.

We adapt the KOTO reconstruction to examine the $6!/(2^3 \times 2!) = 45$ possible sets of pairwise combinations of the six photon hits into 
two pairs that are constrained by Eq.~\eqref{eqn:pimass}, plus one unconstrained pair.
These are labelled by $\alpha = 1,2,\ldots,45$.
For each ``allowed'' set, i.e. one whose two constrained pairs both reconstruct a vertex, 
one computes a weighted average vertex as in Eq.~\eqref{eqn:barzv}.
One again determines the reconstructed photon pair combination by finding the minimal $\chi^2_\alpha$,
using just the two vertices from the two constrained pairs in each combination.
The weighted average vertex of the combination with $\text{min}(\chi^2_\alpha)$ 
is assigned to be the location of the reconstructed $K_L$ vertex, $z_v$, on the $z$ axis,
and we determine the perturbative transverse displacement $\vec{r}_v$ from the center-of-energy $\vec{r}_E$ on the CsIC.

\begin{figure}[tb]
	\includegraphics[width = 0.8\linewidth]{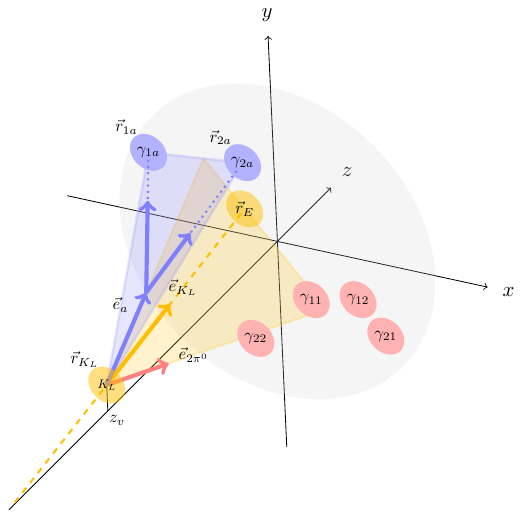}
	\caption{Geometry of reconstructed kinematics for $K_L \to 2\pi^0 a$, applying Eq.~\eqref{eqn:pimass} to only two diphoton pairs (red),
	followed by the selection of Sec.~\ref{sec:kotodetsim} to determine the $K_L$ decay vertex (orange).
	Under the modified selection to determine $\malprec$, 
	we require the reconstructed ALP momentum $p_a$ to be coplanar with that of $K_L$ and $2\pi^0$ (orange plane)
	and also lie in the plane defined by the $a \to \g\g$ photon hits (blue) and the $K_L$ decay vertex (blue plane).}
	\label{fig:recoALP}
\end{figure}

With reference to Fig.~\ref{fig:recoALP}, the direction of the $K_L$ momentum $\vec{e}_{K_L}$ must lie on $\vec{r}_E - \vec{r}_{T1}$, 
the location of the T1 target, which is treated as pointlike  
(see Sec.~\ref{sec:kotodetsim} for the definition of $\vec{r}_E$),
while the summed spatial momentum of the two pions $\vec{p}_{2\pi^0}$ is also known,
and thus the corresponding unit direction $\vec{e}_{2\pi^0}$.
The ALP spatial momentum should be co-planar with $\vec{e}_{K_L}$ and $\vec{e}_{2\pi^0}$,
and it should also be coplanar with the plane formed by the reconstructed $K_L$ vertex at $\vec{r}_{K_L} = \{\vec{r}_v, z_v\}$ 
and the two calorimeter hits for the unconstrained photon pair ($\g_{1a,2a}$ in Fig.~\ref{fig:recoALP})
at $\vec{r}_{ia}$, $i = 1$, $2$, respectively.
Thus the direction of the ALP momentum $\vec{e}_a$ is known via 
\begin{equation}
	\label{eqn:ea} 
	\vec{e}_a \sim \big[(\vec{r}_ {1a} - \vec{r}_{K_L}) \times (\vec{r}_ {2a} - \vec{r}_{K_L}) \big] \times \big[\vec{e}_{K_L} \times \vec{e}_{2\pi^0} \big]\,,
\end{equation}
in which we use the notation `$\vec{e} \sim \vec{k}$' to mean the unit vector $\vec{e} = \vec{k}/|\vec{k}|$.

Defining an ALP decay displacement parameter, $\lambda$, 
then the unit directions of the ALP daughter photon momenta $\vec{e}_{ia}(\lambda) \sim \vec{r}_ {ia} - (\lambda\vec{e}_a + \vec{r}_{K_L})$,
and the momenta $\vec{k}_{ia}(\lambda) = E_{ia}\vec{e}_{ia}(\lambda)$.
At truth level, $\vec{e}_a \sim \vec{k}_{1a} + \vec{k}_{2a}$, 
so the spatial momenta obey the relation $(\vec{k}_{1a} + \vec{k}_{2a})^2 - [(\vec{k}_{1a} + \vec{k}_{2a})\cdot\vec{e}_a]^2=0$.
However, $\vec{e}_a$ is necessarily always imperfectly reconstructed via Eq.~\eqref{eqn:ea} because of finite detector resolution effects.
Instead, we seek a value for $\lambda$, $\lambda_{K_L}$, such that the total invariant mass 
\begin{equation}
	\label{eqn:lamKL}
	\sqrt{\big[p_{2\pi^0} + k_{1a}(\lambda_{K_L}) + k_{2a}(\lambda_{K_L})\big]^2} = m_{K_L}\,,
\end{equation}
whence the reconstructed ALP mass $\malprec = \sqrt{[k_{1a}(\lambda_{K_L}) + k_{2a}(\lambda_{K_L})]^2}$, 
the diphoton invariant mass.
We do not reconstruct events for which Eq.~\eqref{eqn:lamKL} has no solution.
(Alternatively, in the case of no solution one may select $\lambda_{K_L} = 0$, which also appears to work well for reconstructing $m_a$, 
perhaps as zero is an unbiased estimator for the true $\lambda_{K_L}$.)

\subsection{Signal and background estimation}
\label{sec:background}

Unlike in Sec.~\ref{sec:fitlim}, in which we used the measured $\mKLrec$ distribution to set limits,
to illustrate the potential KOTO limits from $\malprec$ 
we must estimate both the $K_L \to 3\pi^0$ contribution as well as the continuum from accidental overlay.
To simulate the $K_L \to 3\pi^0$ contribution, 
we generate a simulated sample with $3\times 10^7$ initial $K_L$'s at the BE,
and pass it through the above-described modified reconstruction,
which we emphasize imposes different requirements versus those for the $\mKLrec$ KOTO data in Sec.~\ref{sec:KOTOsel}.
In particular, for the third diphoton pair, we set the small but finite lifetime $c\tau = c\tau_\pi$ and require Eq.~\eqref{eqn:lamKL} to have a solution.
As before, we assume that the differential decay rate is well approximated as pure phase space.

\begin{figure}[tb]
	\includegraphics[width = \linewidth]{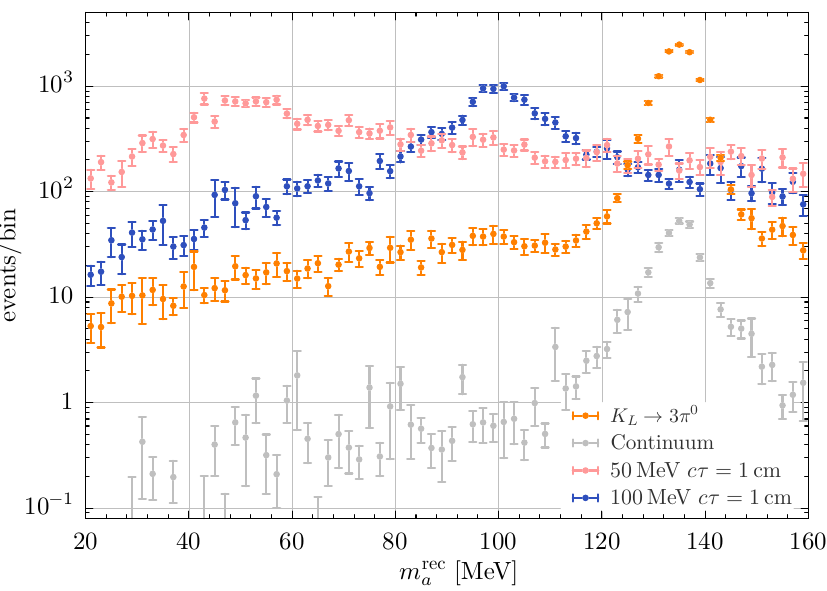}
	\caption{Distributions of the reconstructed ALP mass, $\malprec$, per the selection of Sec.~\ref{sec:bych} 
	and with statistics roughly corresponding to $2 \times 10^{14}$\,POT
	for three cases: $m_a = m_{\pi}$ with $c\tau = c\tau_\pi$ (orange), corresponding to the background from $K_L \to 3\pi^0$; 
	$m_a = 100$\,MeV with $c\tau = 1$\,cm (blue); and $m_a = 50$\,MeV with $c\tau = 1$\,cm (light red).
	Also shown is the contribution the continuum background (light gray).}
	\label{fig:recoALPmass}
\end{figure}

The statistics of a sample of this size, when passed through the selection of Sec.~\ref{sec:kotodetsim}, 
roughly corresponds to the statistics in the KOTO $\mKLrec$ data for $2\times 10^{14}$ POT, 
i.e. $70 \times 10^3$ events.
The acceptance of the modified reconstruction is approximately five times lower, 
such that one obtains a sample with statistics equivalent to $12 \times 10^3$ events.
The resulting $\malprec$ distribution is shown in Fig.~\ref{fig:recoALPmass} (orange).
We also show in Fig.~\ref{fig:recoALPmass} an example distribution for $m_a = 100$\,MeV with $c\tau = 1$\,cm (blue)
and for $m_a = 50$\,MeV with $c\tau = 1$\,cm (light red).
In the latter examples, typical ALP displacement $c\tau\beta\gamma$ can be $\gtrsim 10$\,cm.
For the $\pi^0$ and $100$\,MeV cases, the pseudoscalar mass is faithfully reconstructed,
while for $50$\,MeV the peak starts to degrade.

To estimate the accidental overlay continuum is more challenging.
As an approximate treatment we imagine the continuum can be well approximated by $K_L \to 3\pi^0$, 
but allowing the parent $K_L$ mass to vary uniformly between $3m_\pi^0$ and $600~\text{MeV}$, 
per the range of the $\mKLrec$ KOTO data.
To approximate the relative contributions of $K_L\to 3\pi^0$ events ($B_{3\pi^0}$) 
versus the continuum ($B_{\text{cont}}$) to the total background $B_{\text{tot}}$,
i.e. to determine the fraction $f$ in
\begin{equation}
	\label{eqn:Btot}
    	B_{\text{tot}} = (1-f) B_{3\pi^0}+ f B_{\text{cont}}, 
\end{equation}
we simulate normalized $\mKLrec$ distributions for $B_{\text{cont}}$ and $B_{\text{tot}}$, to be fit to the measured $\mKLrec$ spectrum.
We generate a $B_{\text{cont}}$ sample with $\simeq 5 \times 10^6$ initial $K_L$'s.
Fitting the linear combination~\eqref{eqn:Btot} to the measured $\mKLrec$ spectrum, we obtain $f \simeq 0.03$.
As $\simeq 99\%$ of the $K_L \to 3\pi^0$ contribution lies within an approximately $\pm 16$\,MeV window of the peak,
then the continuum contribution dominates the background outside this window, as expected from Fig.~\ref{fig:KOTOrec}.
The 68\% CL band for the ratio of this MC simulation versus the $\mKLrec$ data is shown in Fig.~\ref{fig:KOTOrec} (bottom panel, light red band),
and exhibits generally good agreement across the entire $\mKLrec$ range.
The corresponding continuum contribution to the $\malprec$ distribution is shown in Fig.~\ref{fig:recoALPmass} (light gray).
It features the same morphology as the $K_L \to 3\pi^0$ background (in orange), 
and can be safely neglected for the purpose of estimating limits.

\subsection{Naive projections}
For the purpose of demonstrating the approximate reach of KOTO, 
we use the purely statistical uncertainties of the combined background $B_{\text{tot}}$ 
to naively approximate the full uncertainties in a KOTO measurement with statistics corresponding to $2\times 10^{14}$ POT,
and treat them as residuals to which a signal hypothesis may be compared.
To obtain naive limits, we generate simulated samples for the $K_L \to 2\pi^0 a$ signal including finite ALP displacements,
with $3 \times 10^6$ initial $K_L$'s, which is again sufficient for subleading statistical uncertainties versus the background.
We consider $m_a = 40, 50, \ldots, 130$~MeV and $c\tau= c\tau_\pi, 1$\,cm and $10$\,cm,
using a 2~MeV binning for the $\malprec$ distribution between 20~MeV and 160~MeV.

\begin{figure}[tb]
    \includegraphics[width=\linewidth]{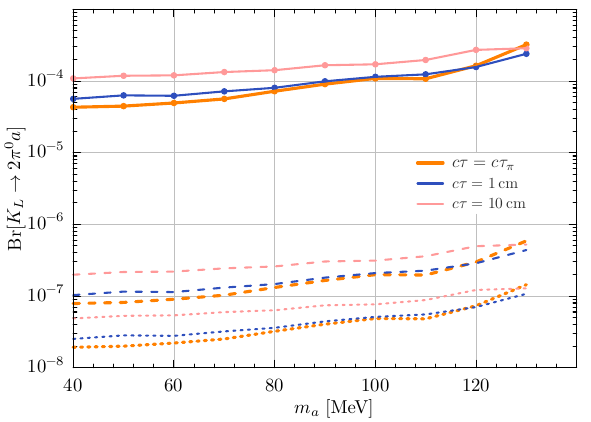}
    \caption{Naive estimates for the 90\% CL limits for $\text{Br}\left[K_L\to2\pi^0a\right]$ with $2\times10^{14}$~POT (solid), 
    $6\times10^{19}$~POT (dashed) and $10^{21}$~POT (dotted) for ALPs with lifetime $c\tau = c\tau_\pi$ (orange), $1$\,cm (blue) and $10$\,cm (magenta).}
    \label{fig:projections}
\end{figure}

Assuming that the signal MC uncertainties remain negligible and that there are no bin-bin correlations,
we set a 90\% CL limit for a given mass and lifetime hypothesis via
$\chi^2(m_a, c\tau) \simeq \sum_i \big[ \rho S_i(m_a, c\tau) /\sigma_{B_{\text{tot}}, i} \big]^2$,
with $\rho$ defined as in Eq.~\eqref{eqn:rhodef}, 
$S_i$ the $i$th signal bin for $K_L \to 2\pi^0 a$, 
and $\sigma_{B_{\text{tot}}, i}$ the background uncertainty for that bin.
Because $\rho \ge 0$ the $\chi^2$ statistic is an $1:1$ admixture of a $0$ and $1$ degree of freedom $\chi^2$, 
as above, the 90\% CL corresponds to $\chi^2(m_a, c\tau) \le 1.64$.
The projected branching ratio limits are shown in Fig.~\ref{fig:projections} for $2\times10^{14}$~POT (solid),
and are similar to the limits obtained at the boundary of the pion chimney.
The dependence on $m_a$ is somewhat weak over the considered mass range,
except within the pion chimney range in which they begin to weaken:
whether future limits set inside the pion chimney via the $\malprec$ spectrum are more powerful than those from $\mKLrec$ 
requires a dedicated study within the experimental framework itself.
We also show limits for $6\times10^{19}$~POT,
roughly corresponding to the size of the 2016-2021 runs,
and the expected future $10^{21}$~POT (dotted), 
in both cases simply scaling by $1/\sqrt{\text{POT}}$.

While one might expect that signals with larger ALP lifetimes should exhibit cleaner separation versus the prompt $3\pi^0$ background,
we see in Fig.~\ref{fig:projections} that the $c\tau = 1$ and $10$\,cm benchmarks have slightly weaker limits
versus the prompt case.
This may in part be because of isolation requirements, that favor ALPs decaying farther from the CsIC.
Similarly, for larger lifetimes one expects more and more ALPs to decay beyond the DV and after the CsIC, 
reducing the acceptance: 
In the long-lifetime regime, e.g. $c\tau=1$\,m, for which the typical displacement is well-beyond the DV, 
the fraction of decays in the DV scales as $1/c\tau$, 
so that the branching fraction limits should correspondingly weaken.
We anticipate that further study of the selection and reconstruction for $\malprec$ within the full experimental framework,
including a study of tight timing requirements,
may be able to achieve improved efficiences for displaced ALPs.

Comparing our estimate of the branching ratio reach in Fig.~\ref{fig:projections} 
to the predicted branching ratios and corresponding lifetimes in the different benchmark scenarios in Fig.~\ref{fig:KOTO_bounds_combined}, 
we observe that a search for $K_L\to2\pi^0a\to6\gamma$ at KOTO could be competitive 
with searches for the two-body decay $K_L\to\pi^0a$ in the mass range $m_a<m_{\pi^0}$,
particularly in the benchmark model shown in the top of Fig.~\ref{fig:KOTO_bounds_combined}. 
In general, however, these neutral kaon searches have less constraining power than those involving the corresponding charged kaon decay, $K^+\to\pi^+a$.

\section{Summary and Conclusions}
\label{sec:conc}
In this work we showed that $K_L \to 2\pi^0a$ decays, with ALP masses within $\mathcal{O}(10)$\,MeV of the neutral pion's,
are able to pass the KOTO selection for the $K_L \to 3\pi^0 \to 6\g$ calibration data with $2\times 10^{14}$ POT.
These decays induce additional single or multipeak structures in the reconstructed $K_L$ mass spectrum, $\mKLrec$,
that are shifted with respect to the peak at $m_{K_L}$.
This allows for the KOTO calibration data to be used in principle to set limits on $\text{Br}[K_L \to 2\pi^0a]$,
for ALP mass ranges that are typically hard to probe by other means: a mass range we have called the pion chimney.

In the absence of priors for SM background contributions to the $\mKLrec$ lineshape, 
we showed that one may nonetheless derive limits on $\text{Br}[K_L \to 2\pi^0a]$
via phenomenological Voigt-style parametrizations of the lineshape profile, 
combined with standard information theoretic techniques and hypothesis testing.
In the ALP mass range  $120 \le m_a \le 150$\,MeV, 
we find branching ratio limits of order $\text{few}\times 10^{-4}$ to $\text{few}\times 10^{-3}$,
which correspond to novel limits in the $m_a$ vs ALP coupling plane for several commonly considered simplified ALP models.
These results should be considered prospective,
in the sense that model dependence effects from our phenomenological lineshape description,
that appear to be at most $\mathcal{O}(1)$,
require quantification from precision simulations of the accidental overlay continuum contributions and other backgrounds.
Further, a fully-fledged analysis within an experimental framework is required to quantify 
and incorporate any correlations in the measured $\mKLrec$ spectrum.

In addition, we explored the potential for KOTO to set branching ratio limits on $\text{Br}[K_L \to 2\pi^0a]$
via a modified selection that incorporates the possibility of ALP displaced decays-in-flight.
This technique can be applied to ALP masses outside the pion chimney.
Naive limits are obtained via the reconstructed diphoton mass spectrum from $a \to \g\g$, i.e. $\malprec$,
that are comparable to the limits obtained via $\mKLrec$ within the chimney. 
These beyond-chimney limits may be competitive with other neutral kaon modes for a larger POT dataset
depending on the NP model,
though they generally cannot be competitive with limits from $K^+ \to \pi^+ X$ searches.
However, extension of either the $\mKLrec$ or $\malprec$-based searches to larger (future) KOTO datasets 
may permit almost the entirety of the pion chimney itself to be probed.

\begin{acknowledgements}
We thank Florian Bernlochner, Vladimir Gligorov, Yuval Grossman, and Simon Knapen for discussions.
The research of RB and SG is supported in part by the U.S. Department of Energy grant number DE-SC0010107. 
This work was performed in part at the Aspen Center for Physics, which is supported by National Science Foundation grant PHY-2210452. 
This research was supported in part by grant NSF PHY-2309135 to the Kavli Institute for Theoretical Physics (KITP).
DJR and CS are supported by the Office of High Energy Physics of the U.S. Department of Energy under contract DE-AC02-05CH11231.
\end{acknowledgements}

\bibliography{chimneyK2PiA.bib}

\end{document}